\documentclass[twocolumn]{aastex701}
\usepackage{natbib,amssymb,amsmath,graphicx,here,gensymb,xcolor,txfonts,hyperref}
\begin{document}

\title{High-contrast L-band Integral Field Spectroscopy of HD~33632~Ab}

\correspondingauthor{Jordan Stone}
\email{jordan@lbti.org}

\author[0000-0003-0454-3718]{Jordan M. Stone}
\email{jordan@lbti.org}
\affiliation{Naval Research Laboratory, 
Remote Sensing Division, 
4555 Overlook Ave SW, 
Washington, DC 20375 USA} 

\author[0000-0002-2314-7289]{Steve Ertel}
\email{sertel@lbti.org}
\affiliation{Department of Astronomy and Steward Observatory,
University of Arizona,
933 N. Cherry Ave,
Tucson, AZ 85721-0065 USA} 
\affiliation{Large Binocular Telescope Observatory,
University of Arizona,
933 N. Cherry Ave,
Tucson, AZ 85721-0065 USA} 

\author[0000-0002-7129-3002]{Travis Barman}
\email{barman@arizona.edu}
\affiliation{Lunar and Planetary Laboratory, 
The University of Arizona, 
1629 E. Univ. Blvd., 
Tucson, AZ 85721 USA}

\author[0000-0001-6098-3924]{Andrew J. I. Skemer}
\email{askemer@ucsc.edu}
\affiliation{Department of Astronomy and Astrophysics, 
University of California, Santa Cruz, 
1156 High St, 
Santa Cruz, CA 95064, USA}

\author[0000-0002-0834-6140]{Jarron M. Leisenring}
\email{jarronl@arizona.edu}
\affiliation{Steward Observatory,
University of Arizona,
933 N. Cherry Ave,
Tucson, AZ 85721-0065 USA} 

\author[0000-0002-1954-4564]{Philip M. Hinz}
\email{phinz@ucsc.edu}
\affiliation{Department of Astronomy and Astrophysics, 
University of California, Santa Cruz, 
1156 High St, 
Santa Cruz, CA 95064, USA}

\author[0000-0001-6567-627X]{Charles E.\ Woodward}
\email{woodw024@umn.edu}
\affiliation{Minnesota Institute for Astrophysics, 
School of Physics and Astronomy, 
116 Church Street, S.E., 
University of Minnesota, 
Minneapolis, MN 55455, USA}

\author{Michael F. Skrutskie}
\email{mfs4n@virginia.edu}
\affiliation{Department of Astronomy, University of Virginia, Charlottesville,
VA 22904, USA}

\begin{abstract} 

We present LBTI/ALES $3.07-4.08\micron$ spectroscopic observations of
HD~33632~Ab, a $\sim53~M_{Jup}$ directly imaged companion to an F8 star.
Spectroscopic measurements of HD~33632~Ab now span $1-4~\micron$, and we
perform the first spectroscopic analysis covering this full range. The data are
compared to isolated brown dwarf template spectra, indicating that HD~33632~Ab
is similar to L8/9 field brown dwarfs. Synthetic atmosphere model spectra from
multiple model families are fit, with cloudy models providing the best fits,
consistent with expectations for an L-dwarf.  Evolutionary model predictions
for the bulk properties of HD~33632~Ab are highly constrained by the precise
dynamical mass found for the object. In particular, predictions for surface
gravity are narrowly peaked, $\log(g)=5.21\pm0.05$, and not dependent on the
effects of clouds or cloud dispersion. We find significant tension between the
surface gravities and object radii inferred from atmosphere model fits and
those predicted by evolutionary models. We conclude with a comparison to the
spectra of the HR~8799~c, d, and e, and emphasize the case that HD~33632~Ab,
and other L/T transition directly imaged companions with constrained masses,
will serve an essential role in understanding the complex physical processes
governing the appearance of clouds in planetary atmospheres.

\end{abstract}


\section{Introduction} 
The \textit{Gaia} mission \citep{GaiaCollaboration2016} is transforming the way
exoplanet direct imaging science is done. The most obvious impact is the
increased efficiency of direct imaging searches, with several newly discovered
high-contrast substellar companions arising from target lists composed of stars
showing astrometric acceleration \citep[e.g., HD~33632~Ab, HIP~21152~B,
HD~99770~b, AF~Lep~b, HIP~39017~b ][]{Currie2020, Bonavita2022, Currie2023,
Mesa2023, Tobin2024}. Recent results show substellar companion detection rates
as high as $16\%$ when using \textit{Gaia}-assisted targeted searches
compared to $\lesssim1\%$ for blind searches \citep{Bonavita2022,
Stone2018}.

Beyond detection, \textit{Gaia}-assisted direct imaging discoveries are poised
for detailed atmospheric characterization because dynamical mass constraints
can be harnessed to guide atmosphere modeling, a fact long appreciated by the
transiting planet community. \citet{Zhang2023} demonstrated the power of this
approach, confidently showing that the directly imaged planet AF~Lep~b is metal
enriched compared to its host star. It is now clear that constraining the
fundamental parameters of substellar objects is essential for enabling the most
detailed atmospheric characterization and for achieving the goals of the
exoplanet community to understand cloud physics, atmospheric dynamics,
atmospheric composition, planet formation and evolution
\citep[see][]{Calamari2024}.

Here we present new spectroscopic measurements of HD~33632~Ab, a brown dwarf
companion discovered $\approx0\farcs75$ in projected separation from its host
star by \citet{Currie2020} in a campaign to target astrometric accelerators.
\citet{ElMorsy2025} combine astrometric acceleration measurements with radial
velocity and projected separation measurements to constrain the system orbit
and derive a secondary mass of $52.8^{+2.6}_{-2.4}~M_{\mathrm{Jup}}$. 

The host star, HD~33632~Aa, is an F8V type star at 26.38~pc with an estimated
age of $1.5^{+3.0}_{-0.7}$~Gyr \citep{Gray2003, GaiaCollaboration2021,
Currie2020}. \citet{Rice2020} analyze Keck/HIRES observations of HD~33632~Aa
and constrain stellar effective temperature, surface gravity and the abundance
of several elemental species.

Multiple studies have analyzed infrared spectroscopic measurements of
HD~33632~Ab to infer its atmospheric properties. \citet{Currie2020} compare
near-infrared (NIR) low resolution, $R\sim20$, Subaru/Charis observations
covering the J-, H-, and K-bands to brown dwarf spectral templates, finding
a best fit of L9.5.  Follow-up observations targeting the H- and K-bands at
higher spectral resolution with Subaru/Charis are reported by \citet{Gibbs2024}
and \citet{ElMorsy2025}. \citet{ElMorsy2025} combine these data with L-band
photometry from \citet{Currie2020} to perform atmosphere model fits, finding
best-fit models consistent with evolutionary model predictions. \citet{Hsu2024}
use the Keck/KPIC module to direct light from HD~33632~Ab into the NIRSPEC
instrument for $R\sim35000$ spectroscopy within the K-band. These data are
harnessed to constrain the rotation velocity of the brown dwarf and to measure
chemical abundances, which include a non-detection of methane. Each of the
previous studies has made use of a combination of the several spectral ranges
covered, including JHK, HKL, only H-band or only K-band. Many previous studies
demonstrate that using broad wavelength coverage breaks degeneracies in spectroscopic
analysis investigating the atmospheres of brown dwarfs
\citep[see][]{Stephens2009, Sorahana2014} and directly imaged companions
\citep[see][]{Skemer2016, Stone2020, Doelman2022}. The L-band, covering
$\sim2.9-4.1\micron$, is particularly sensitive to the effects of clouds and
non equilibrium carbon chemistry.  Here we present the first analysis of
spectroscopic measurements spanning J-, H-, K-, and L-band.

\section{Observations and data reduction}\label{sec:observations_and_data_reduction} 

We observed HD~33632~Ab on UT 2021 Feb 28 using the Large Binocular Telescope
(LBT). Our setup directed light from the SX (or left side) 8.4~m LBT aperture
to the Arizona Lenslets for Exoplanet Spectroscopy (ALES) integral field
spectrograph \citep[IFS,][]{Skemer2018} mode of the L- and M-band Infrared Camera
(LMIRCam) instrument \citep{Leisenring2012}, which operates within the
framework of the Large Binocular Telescope Interferometer \citep[LBTI,
][]{Hinz2016, Ertel2020}.  ALES provides $R\sim40$ spectra from 2.9 to 4.2
$\micron$ over a $\sim2\farcs2$ field of view. Our observational setup made use
of an 8x refractive magnifier and a lenslet array with $500~\micron $ lenses,
yielding 34.5 milliarcseconds spatial sampling \citep{Stone2022}. No
coronagraph or apodizing optics are used. Light from the DX (or right side)
8.4~m LBT aperture was directed into the PEPSI instrument
  \citep{Strassmeier2015} to enable detailed spectral analysis of the host
star, including inference of the system bulk composition. Those data will be
reported in a future article.

Observations began just after HD~33632~A transited the meridian. We tracked the
system for a total of 140 minutes accumulating $39^{\degree}$ of parallactic
angle rotation.  Atmospheric seeing at visible wavelengths varied from
$1\farcs1$ to $1\farcs5$. The adaptive optics system was run correcting 400
modes at 1.4 kHz loop speed.

A nodding sequence was used to track variable background emission, collecting
300 1.48s frames on-source and then 300 1.48s frames off source each cycle for
a total of 5 cycles, totaling of 33.9 m of on-source exposure time.

\subsection{Raw Frame Preprocessing} 
Our 1.48s integrations include three non-destructive reads each. To remove
detector reset noise and bias drifts we subtract the first read from the last
read of each ramp (the central read is disregarded in our analysis).

We subtract an estimate of the background emission from each on-source frame.
To do this, we identify the 200 closest-in-time off-source frames and median
combine them. After background subtraction, residual channel offsets are
removed using the median of the overscan pixels in the lowest four rows of each
channel. Vertical overscan columns are median combined and smoothed with
a Savitsky-Golay filter using a 31-pixel window length and a polynomial order
of three \citep{Press1990}. The result is subtracted from each column.

Bad pixels are identified using a set of dark frames, modeling the distribution
of pixel values as a Gaussian, and finding those pixels with values having less
than 1$\%$ probability of arising by chance given the more than 4 million
pixels in the array. We take the resulting bad pixel mask and correct each bad
pixel in our background and overscan corrected images by replacing them with
the median of their four closest neighbors.

\subsection{Creating Spectral Cubes} 

We transform our two-dimensional images, composed of a grid of short
interleaved spectral traces, into data cubes with two spatial dimensions and
one wavelength calibrated spectral dimension following the approach described
in \citet{Stone2018}, and \citet{Stone2022}. The ALES pipeline uses the optimal
extraction algorithm \citep{Horne1986} on each microspectrum sequentially,
which can be more susceptible to mircorspectra cross talk than $\chi^2$
extraction technique used in the Charis pipeline \citep{Brandt2017}.

In order to reduce computer processing time in subsequent analysis steps, we
bin our data cubes in the spectral dimension to reduce oversampling. We
assume a constant spectral resolution across the band of $R\sim40$ \citep{Stone2022}
and resample the spectrum at each spatial location using the
SpectRes package \citep{Carnall2021} to yield three evenly spaced samples per
resolution element.

\subsection{High Contrast Image Processing} \label{sec:classical_adi} 

\begin{figure} 
\includegraphics[width = \linewidth]{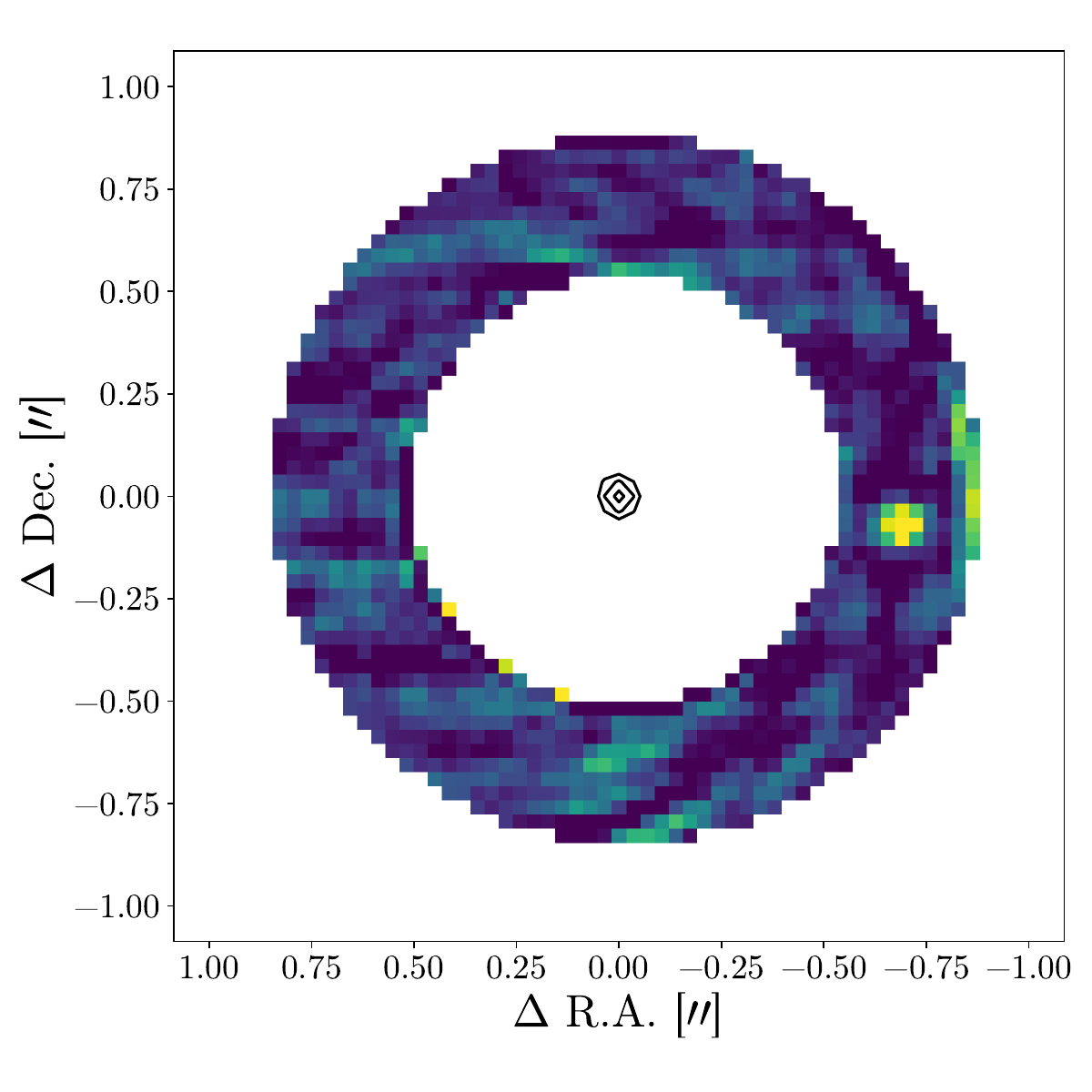} 
\caption{The wavelength stacked image of HD~33632~Ab taken with LMIRCam+ALES.
The annular region shows the pixels processed with our PCA-based ADI routine.
White regions have been masked for display. Black contours near the origin
indicate the position of the unsaturated host star.}
\label{fig:finalImage} 
\end{figure}

The orientation of our images rotates with the parallactic angle, decreasing
from $134\arcdeg$ to $85\arcdeg$ over the course of our observation.  We use
angular differential imaging \citep[ADI,][]{Marois2006}  to processes our
frames implementing a principal component analysis \citep[PCA,][]{Amara2012,
Soummer2012} approach. We treat each wavelength slice individually focusing on
an annulus centered on the host star with radius equal to the companion
separation and width $\sim5\lambda/D$ (12 pixels).  For each frame we create
a point spread function (PSF) library using all other frames excluding those
with the companion position within 2 pixels.  After host-star PSF subtraction
at each wavelength, images are rotated to orient North up and median combined
to produce a final reduced image cube.  We tested reductions using 3, 10, and
30 principal components when subtracting the host star; using 3 principal
components provided the best signal-to-noise ratio of the companion in the
final stacked image. Figure \ref{fig:finalImage} shows the image that results
from summing over the wavelength dimension of our final data cube. It is
noteworthy that 3 principal components optimizes the reduction given a library
of 1500 frames.  This results from low illumination of the speckle halo
(compared to the thermal background) at the separation of HD~33632~Ab in the
ALES data. As we note below, background dominated noise also affects the shape
of the spectral covariance matrix.

The signal to noise ratio for HD~33632~Ab is estimated in each
wavelength slice using a PSF-weighted mean of pixels within an aperture of
radius corresponding to $\lambda$/$D$. Specifically, we take signal as the flux
in an aperture centered on the brightest pixel in the wavelength collapsed
cube, and noise is estimated as the standard deviation of PSF-weighted flux
measurements in non-overlapping apertures in a ring about the primary star with
the same projected separation, avoiding regions of significant detector
crosstalk \citep[see][]{Doelman2022}. Signal-to-noise maps created using the
same approach are presented in the appendix.

\subsection{Spectral Extraction} \label{sec:spectralExtraction} 

High-contrast image processing algorithms can bias companion photometry due to
over subtraction. To accurately extract the flux of HD~33632~Ab we use
a forward modeling approach, subtracting a model of the companion
from each frame prior to running PCA PSF optimization. The host star did not
saturate in any of our frames, so we use it as a time-resolved PSF reference to
model the companion. This provides photometric measurements robust to variable
atmospheric throughput and changes in the PSF shape due to instrumental flexure
and variable AO-performance.  We fit for a scale factor and position of the
companion, using the sum of squared residuals ($\chi^2$) in a $3\times3$ pixel
region centered on the object as the figure of merit.  

Brute force forward modeling is computationally expensive with 1500 frames.
Therefore, we choose to ignore 19 wavelength slices between $3.07~\micron$ and
$3.5~\micron$ where signal-to-noise as determined using aperture photometry was
below 2.  For slices with signal-to-noise $>2$, we use a grid-fitting approach
to determine the relative brightness of the companion compared to its host
star.  First, we calculate $\chi^2$ for 10 scale factors logarithmically spaced
between $10^{-4}$ and $10^{-3}$ with the companion position fixed at the
location of the brightest pixel in the median image.  Next, we fit a parabola
to the $\chi^2$ versus scale factor data and identify the scale factor that
minimizes the parabola.  This scale factor provides a good-guess starting place
for our simultaneous fit of position offset and scale factor, described next.

\begin{figure} \includegraphics[width = \linewidth]{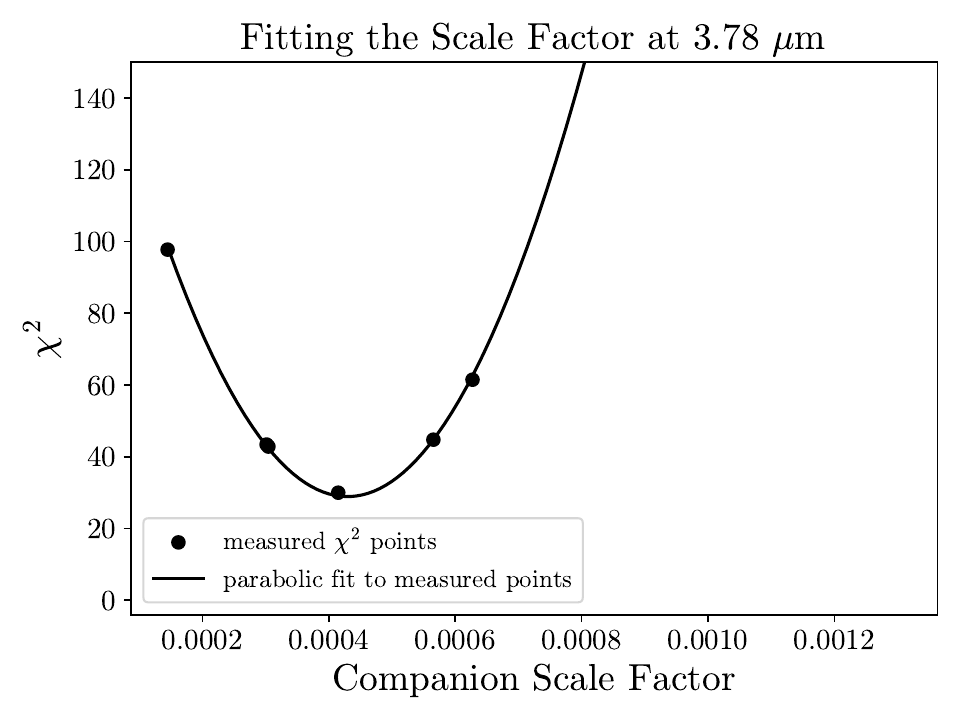}
\caption{Scale factors quantifying the flux ratio between the companion and
host star are determined for each wavelength by minimizing a parabola fit to
the results of a grid-fitting algorithm.} \label{fig:SpecExtractChi}
\end{figure}

For each of 9 positions on a square grid with quarter-pixel steps, the best-fit
scale factor is determined by minimizing a parabola fitted to seven scale
factor versus $\chi^2$ points, four logarithmically spaced between
$\frac{1}{3}$ and $3$ times the good guess scale factor and three
logarithmically spaced between 0.7 and 1.3 times good guess scale factor.
Figure \ref{fig:SpecExtractChi} illustrates our approach for the $3.78~\micron$
slice. Minimum $\chi^2$ values for each wavelength are found with a position
offset one quarter pixel to the south of the brightest pixel in the collapsed
cube.

\begin{figure} \includegraphics[width = \linewidth]{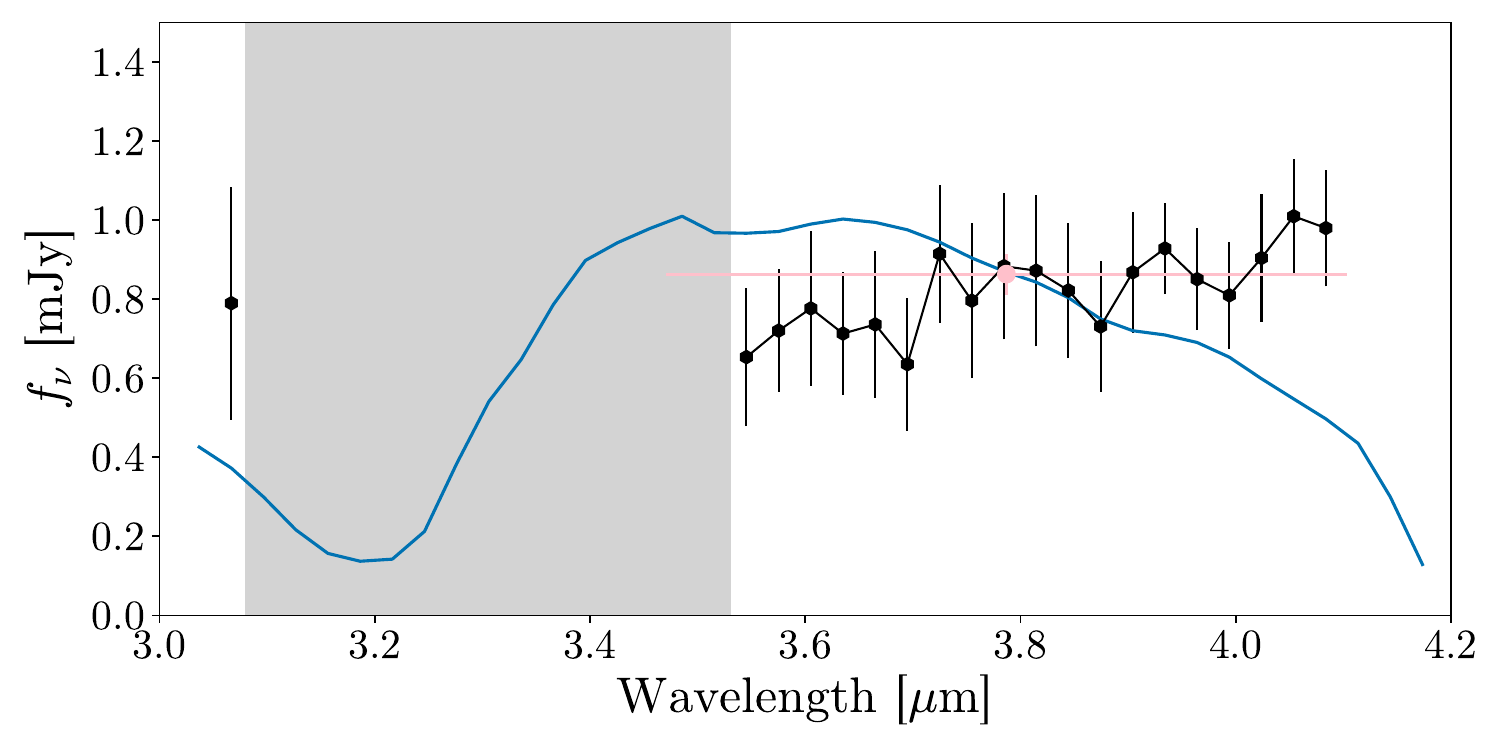}
\caption{The flux calibrated ALES spectrum and uncertainties of HD~33632~Ab
(black markers) compared to the \citet{Currie2020} Keck/NIRC2 L-band
photometric measurement, shown with a pink marker with horizontal errorbars
indicating the filter bandwidth and vertical errorbars indicating photmetric
uncertainty. We indicate with a gray swath a region in which we did not recover
the companion with signal-to-noise ratio greater than 2. The blue curve is
a normalized, uncalibrated spectrum of the central star, highlighting strong
telluric absorption near $3.2~\micron$.} \label{fig:ALESnirc2} \end{figure}

We perform flux calibration by multiplying our scale factors by a model
spectrum for the host star, smoothed to R=40 and sampled and binned in the same
way as our ALES spectral cubes. We use a BT-Settl \citep{Baraffe2015} model
with $T_{\mathrm{eff}}=6100$~K, $\log(g)=4.5$ \citep{Rice2020}, scaled to
provide a best fit to WISE W1, W2, W3, and W4 photometry \citep{Cutri2012}.
Our flux uncertainty is derived using our signal-to-noise measurements
described above. Figure \ref{fig:ALESnirc2} presents our final flux calibrated
spectrum and uncertainties and compares our results to the Keck/NIRC2
photometric point reported by \citet{ElMorsy2025}. The spectral flux
calibration is consistent with the earlier photometry. We tabulate the
calibrated spectrum Table \ref{table:spectrum}.

Figure \ref{fig:ALESnirc2} also shows a normalized spectrum of HD~33632~A
calculated using aperture photometry using a 3-pixel diameter aperture centered
on the brightest pixel seen when the wavelength dimension is median combined.
This spectrum is uncalibrated and its shape should include the intrinsic shape
of the stellar spectrum (nearly Rayleigh-Jeans at these wavelengths) as well as
features arising from chromatic system throughput, including telluric
absorption. The most dramatic feature is a strong telluric water-ice absorption
band near $3.2~\micron$. However, this band does not span the full
$3.07~\micron - 3.5~\micron$ range of wavelengths where the companion is not
detected above a signal-to-noise ratio of 2.  Evidently, low throughput cannot
completely explain our lack of detections in this band, suggesting low
intrinsic flux from the companion at these wavelengths. 

Spectral covariance in the calibrated spectrum is estimated using the approach
suggested by \citet{Greco2016}. Adhering to the notation of that work, our
best-fit model covariance matrix has $A_{\rho}=0.067$, $\sigma_{\rho}=10125$,
$A_{\lambda}=0.37$, $\sigma_{\lambda}=0.039$, and $A_{\delta}=0.56$. The large
value of $\sigma_{\rho}$ results from residuals being background dominated,
rather than speckle dominated, at the separation of HD~33632~Ab. 

\begin{deluxetable}{lll}
\tabletypesize{\footnotesize}
\tablecolumns{4}
\tablewidth{0pt}
\tablecaption{ALES Spectrum of HD~33632~Ab\label{table:spectrum}}
\tablehead{
    \colhead{$\lambda$}& 
    \colhead{$F_{\nu}$}&
    \colhead{$\delta F_{\nu}$}\\
    \colhead{$\micron$}&
    \colhead{mJy}&
    \colhead{mJy}
    }
\startdata
3.067 & 0.790 & 0.295\\
3.545 & 0.653 & 0.174\\
3.575 & 0.720 & 0.155\\
3.605 & 0.777 & 0.196\\
3.635 & 0.713 & 0.155\\
3.665 & 0.736 & 0.186\\
3.695 & 0.635 & 0.168\\
3.725 & 0.915 & 0.175\\
3.755 & 0.796 & 0.196\\
3.785 & 0.883 & 0.185\\
3.815 & 0.872 & 0.191\\
3.845 & 0.822 & 0.171\\
3.874 & 0.731 & 0.165\\
3.904 & 0.868 & 0.154\\
3.934 & 0.928 & 0.115\\
3.964 & 0.851 & 0.129\\
3.994 & 0.810 & 0.135\\
4.024 & 0.904 & 0.161\\
4.054 & 1.010 & 0.145\\
4.084 & 0.980 & 0.147
\enddata 
\end{deluxetable}

\section{Spectral Analysis}
\subsection{Compiling low-resolution data} 
For our spectroscopic analysis, we combine the low spectral resolution integral
field spectrograph observations of HD~33632~Ab, forming an R$\sim20-65$
spectrum spanning $1.16-4.1~\mu\mathrm{m}$ with gaps in regions of strong
telluric absorption.  For the J-, H-, and K-bands we use published
Subaru/Charis data, preferring the available higher-resolution (R$\sim65$) mode
observations for the H-, and K-bands \citep{Currie2020, ElMorsy2025}. For the
combined Charis+ALES data we create a block diagonal covariance matrix taking
the J-band errors to be diagonal \citep{Currie2020} and using the published H-,
and K-band covariance matrices \citep{ElMorsy2025} together with the L-band
covariance matrix from the current work.

\subsection{Fits to Brown Dwarf Template Spectra}\label{sec:Templates}

We fit the combined Charis+ALES spectrum of HD~33632~Ab to a library of brown
dwarf spectra.  The template library is composed of IRTF/Spex and Subaru/IRCS
observations of brown dwarfs. Objects are selected for having measurements
spanning $\sim1-4.12~\mu\mathrm{m}$ reported in \citet{Cushing2005} and/or
\citet{Rayner2009}. We also include the JWST NIRSpec measurements of VHS
J1256-1257B from \citet{Miles2023}. Library objects are summarized in Table
\ref{TemplateTable}.

To prepare for fitting, template spectra are smoothed and interpolated to
match Charis and ALES spectral resolution and wavelength sampling. Since the
errorbars associated with the Charis and ALES high-contrast data are much
larger than the Spex, IRCS, and NIRSpec data, we do not propagate template
uncertainties.

We perform a $\chi^2$ minimization to determine the library spectrum that most
closely resembles the observations of HD~33632~Ab, using 

\begin{equation} 
\chi^{2}(\tau,f)) = \left((fx(\tau)-\mu)^{T}\Sigma^{-1}((fx(\tau) - \mu)\right),
\label{tempEQ}
\end{equation} 
where $x(\tau)$ represents the template spectrum for the object with spectral
type $\tau$, $f$ is a scale factor, $\mu$ is the observed spectrum of
HD~33632~Ab and $\Sigma$ is the block-diagonal covariance matrix for the IFS
data. For each object we varied $f$ to find the optimal fit at each spectral
type. The results of our fit are presented in Table \ref{TemplateTable}.

\begin{figure*} 
\includegraphics[width = \linewidth]{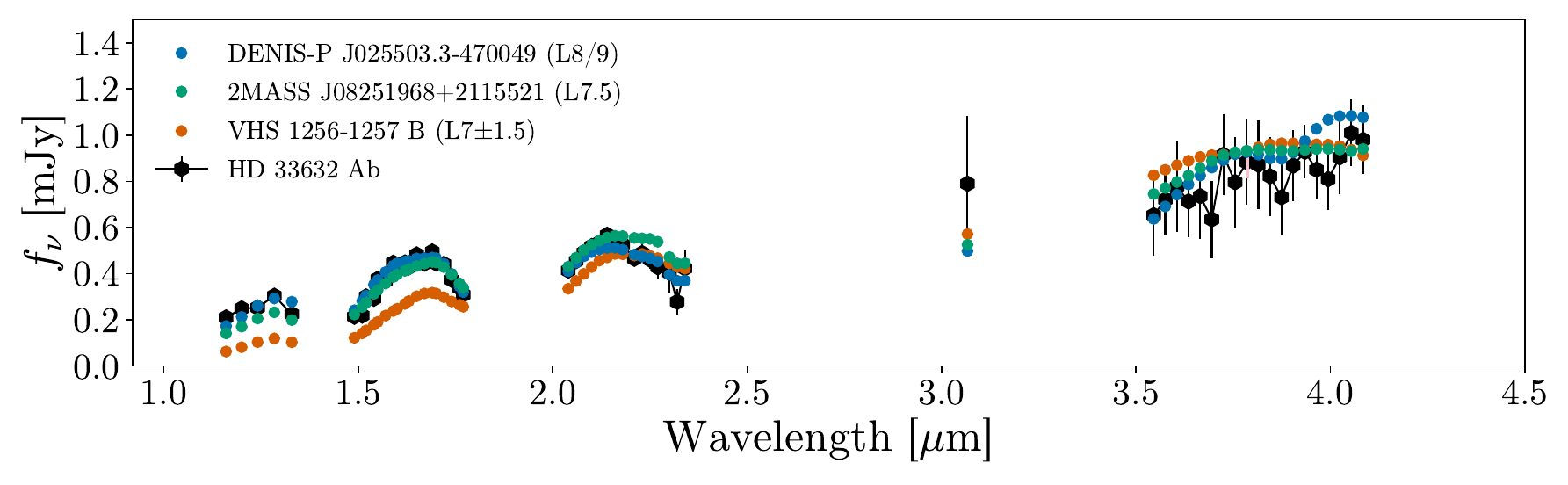} 
\caption{Template brown dwarf spectra compared to HD~33632~Ab.}
\label{fig:TemplatePlot} 
\end{figure*}

The best fit template spectrum is DENIS-P J025503.3-470049, with spectral type
$\sim$L8/9 \citep{Cushing2005, Burgasser2006}. We show the template spectrum
over plotted on the spectral energy distribution (SED) of HD~33632~Ab in Figure
\ref{fig:TemplatePlot}. The template provides a qualitatively good fit to
HD~33632~Ab, especially the J-L, and J-H color. In Figure
\ref{fig:TemplatePlot} we also show the second-best template (the L7.5-type)
and the spectrum of the low-gravity object VHS 1256-1257 B (L7$\pm1.5$).  VHS
1256-1257 B is a poor match to the HD~33632~Ab spectrum, low-gravity resulting
in a redder slope than seen for our target.

Fit quality can be evaluated with the reduced $\chi^2$ metric. Using 56 degrees
of freedom the best fit template has $\chi^{2}_{r}=3.9$, suggesting an
imperfect but reasonable match. We also evaluate the quality of the fit by
comparing the derived scale factor to the expected scale factor. Assuming the
template and HD~33632~Ab have the same effective temperature and radius, the
scale factor should be the ratio of the distances squared. In Table
\ref{TemplateTable} we show the expected and recovered scale factors for each
object and report the difference in units of the uncertainty found by
propagating the reported parallax errors. In Table \ref{TemplateTable} objects
with the least tension are those with the largest parallax uncertainty,
suggesting insufficient precision to constrain the scale factor in these cases.
For other objects, significant tension with the expected scale factor suggests
either that the radii are dissimilar and/or that the effective temperature is
dissimilar. Among the objects with parallax measured with better than 0.5
milliarcsecond precision, DENIS-P J025503.3-470049 has the least tension with
expectations. Systematic issues relating to absolute calibration of the spectra
arising from 3 different instruments may explain the residual discrepancy with
expectation.

\begin{deluxetable*}{lcrrrrr}
\tabletypesize{\footnotesize}
\tablecolumns{4}
\tablewidth{0pt}
\tablecaption{Brown Dwarf Template Library\label{TemplateTable}}
\tablehead{
    \colhead{Obj. Name}& 
    \colhead{Spectral Type\tablenotemark{\footnotesize{a}}}&
    \colhead{$\chi^{2}$}&
    \colhead{parallax\tablenotemark{\footnotesize{b}}}&
    \colhead{$f$\tablenotemark{\footnotesize{c}}}&
    \colhead{expected sf.\tablenotemark{\footnotesize{d}}}&
    \colhead{N$\sigma$\tablenotemark{\footnotesize{e}}}
    }
\startdata
 GJ 1111                  &  M6.5 & 431  &  $279.25 \pm 0.06$ & 0.0004 & 0.018 & 702.815 \\
 GJ 644 C                 &  M7   & 406  &  $153.97 \pm 0.06$ & 0.0019 & 0.061 & 615.836 \\
 LP 412-31                &  M8   & 331  &  $ 68.27 \pm 0.08$ & 0.0107 & 0.308 & 361.683 \\
 DENIS J104814.6-395606   &  M9V  & 360  &  $247.22 \pm 0.05$ & 0.0015 & 0.024 & 648.897 \\
 BRI B0021-0214           &  M9.5 & 234  &  $ 80.34 \pm 0.19$ & 0.0112 & 0.222 & 194.899 \\
 2MASS J14392836+1929149  &  L1   & 242  &  $ 70.56 \pm 0.18$ & 0.0267 & 0.288 & 170.356 \\
 Kelu 1                   &  L2   & 155  &  $ 49.05 \pm 0.73$ & 0.0373 & 0.597 & 31.687  \\
 2MASS J15065441+1321060  &  L3   & 148  &  $ 85.43 \pm 0.19$ & 0.0366 & 0.197 & 174.515 \\
 2MASS J00361617+1821104  &  L3.5 & 150  &  $114.47 \pm 0.14$ & 0.0181 & 0.110 & 299.883 \\
 2MASS J22244381-0158521  &  L4.5 & 305  &  $ 86.15 \pm 0.41$ & 0.0483 & 0.193 & 77.989  \\
 2MASS J15074769-1627386  &  L5   & 114  &  $134.95 \pm 0.26$ & 0.0236 & 0.079 & 170.433 \\
 VHS 1256-1257 B          &  L7   & 293  &   $45    \pm 2.4$  & 0.589  & 0.709 & 1.59 \\
 2MASS J08251968+2115521  &  L7.5 & 97   &  $ 92.60 \pm 0.82$ & 0.1290 & 0.167 & 12.888  \\
 DENIS-P J025503.3-470049 &  L8   & 65   &  $205.40 \pm 0.18$ & 0.0303 & 0.034 & 48.591  \\
 SDSS J125453.90-012247.4 &  T2   & 266  &  $ 74.18 \pm 2.31$ & 0.1829 & 0.261 & 4.812   \\
 2MASS J05591915-1404489  &  T5   & 1025 &  $ 95.27 \pm 0.72$ & 0.0803 & 0.158 & 32.561  
\enddata 
\tablenotetext{a}{Spectral types are taken from \citet{Cushing2005} or
\citet{Rayner2009}. The spectral type of VHS 1256-1257 B is taken from
\citet{Gauza2015}.}
\tablenotetext{b}{Parallax values are reported in milliarcseconds and are taken
from $Gaia$ data releases \citep{GaiaCollaboration2021}. The parallax for VHS
1256-1257 B is taken from
\citet{Dupuy2020}}
\tablenotetext{c}{
The best-fit scale factor from Equation \ref{tempEQ}}
\tablenotetext{d}{The expected scale factor accounts for the different parallax
measurements for HD~33632 and the template dwarfs. It assumes a good spectral
match and similar object radii}
\tablenotetext{e}{The difference between the fitted scale parameter and the
expected scale parameter in units of the propagated uncertainty.}
\end{deluxetable*}

\subsection{Fitting Synthetic Atmosphere Models}\label{sec:Results}

We compare the low-resolution $\sim1-4.12~\mu\mathrm{m}$ spectrum of
HD~33632~Ab to synthetic spectra from four grids of model atmospheres. These
grids include three with clouds and one cloud free. The cloudy grids we use are
Exo-REM, which includes a microphysics-based cloud exposing no adjustable cloud
parameters for model fitting \citep{Baudino2015, Charnay2018, Blain2021},
Sonora Diamondback, which includes a single parameter modifying cloud structure
\citep[fsed,][]{Morley2024}, and the Barman/Brock models, which provide two
parameters to adjust the effects of clouds \citep[$P_{c}$,
$a_{0}$,][]{Barman2011b, Barman2015, Brock2021}.  The cloud free model grid we
use is Sonora Elf Owl, which includes parameters for adjusting both composition
and vertical mixing \citep[nonequilibrium chemsitry,][]{Mukherjee2024}.  Model
grid parameter ranges are summarized in Table \ref{table:grids}.

\begin{deluxetable}{ll}
\tabletypesize{\footnotesize}
\tablecolumns{2}
\tablewidth{0pt}
\tablecaption{Synthetic Atmosphere Model Grid Parameter Ranges\label{table:grids}}
\tablehead{
    \colhead{Parameter}& 
    \colhead{Range}}
\startdata
\cutinhead{Sonora Elf Owl}
$T_{\mathrm{eff}}$               & 1300 K -- 2200 K, 100 K steps \\
$\log(g)$                        & 3.25 - 5.5, 0.25 dex steps \\
$\log(K_{zz})$                   & 2, 4, 7, 8, 9 \\
$\log(\frac{z}{z_{\odot}})$      & -1.0, -0.5, 0.0, 0.5, 0.7, 1.0 \\
$C/O$                            & 0.5 -- 2.5, 0.5 steps \\
\cutinhead{Sonora Diamondback}
$T_{\mathrm{eff}}$               & 900 K -- 2400 K, 100 K steps \\
$\log(g)$                        & 3 -- 5.5, 0.25 dex steps \\
$\log(\frac{z}{z_{\odot}})$      & -0.5 -- 0.5, 0.5 dex steps \\
$f_{\mathrm{sed}}$               & 1, 2, 3, 4, 8, nc \\
\cutinhead{Exo-REM}
$T_{\mathrm{eff}}$               & 400 K -- 2000 K, 50 K steps \\
$\log(g)$                        & 3.00 - 5.0, 0.5 dex steps \\
$\log_{10}(\frac{z}{z_{\odot}})$ & -0.5 -- 2, 0.5 dex steps \\
$C/O$                            & 0.1 -- 0.8, 0.05 steps \\
\cutinhead{Barman/Brock}
$T_{\mathrm{eff}}$               & 800 K -- 1600 K, 100 K steps \\
$\log(g)$                        & 4.75, 5.0, 5.5 \\
$P_{c}$                          & 10 bar -- 30 bar, 10 bar steps \\
$a_{0}$                          & 0.25 $\micron$ -- 0.5 $\micron$, 0.25 $\micron$ steps
\enddata 
\end{deluxetable}

Models are smoothed and sampled to match the different IFS instrument modes.
We interpolate the Barman/Brock grid to 20 K temperature sampling, and 0.1 dex
surface gravity sampling using the same procedure as in \citet{Stone2020}. We
chose not to interpolate either Sonora grid.  For Diamondback, many of the
models are poorly converged \citep{Morley2024} and we opt to avoid
interpolating in this case. For Elf Owl, analysis of an interpolated grid was
too computationally expensive given the high-dimensionality of the models. This
concern also applies to the Exo-REM models.

Best-fit models are determined by maximizing the logarithm of the posterior
function,
\begin{equation}\label{eqn:Fit}
\log{\mathcal{P}} = -0.5\chi^{2} + log(\mathcal{G}) + log(\frac{1}{R})
\end{equation}
Where $\chi^2$ is calculated as 
\begin{equation} 
\chi^{2}(\boldsymbol{\tau},R,\varpi)) = 
\left(((R\varpi)^{2}x(\boldsymbol{\tau})-\mu)^{T}\Sigma^{-1}(((R\varpi)^{2}x(\boldsymbol{\tau}) - \mu)\right), 
\end{equation} 
similar to Equation \ref{tempEQ} with $\mu$ the measured spectrum of
HD~33632~Ab, $\Sigma$ the data covariance matrix, and $x(\boldsymbol{\tau})$
the model spectrum with parameters $\boldsymbol{\tau}$, including effective
temperature, surface gravity, and other model-specific composition and cloud
structure variables. Here we replace the scale factor in Equation \ref{tempEQ}
with the square of the product of $R$, the object radius, and $\varpi$ the
object parallax. The prior, $\mathcal{G}$, incorporates the \textit{Gaia}
parallax measurement and uncertainty $37.8953\pm0.0262$ milliarcseconds
\citep{GaiaCollaboration2021},
\begin{equation}
\mathcal{G}(\varpi) = 
\frac{1}{\sqrt{2\pi(0.0262)^2}}e^{-\frac{(\varpi-37.8953)^2}{2(0.0262)^2}}.
\end{equation}

The last term in Equation \ref{eqn:Fit} represents a log-uniform prior on the
object radius. At this stage, we do not explicitly apply priors to model
parameters, resulting in uniform priors for linear variables
($T_{\mathrm{eff}}$, C/O) and log-uniform priors on logarithmic variables
($\log(g)$, $\log{K_{zz}}$, $\log(\frac{z}{z_{\odot}})$).

We employ a brute-force, grid-fitting method testing each model in each grid
using 266 steps in object radius from 0.5 $R_{\mathrm{Jup}}$ to 2.5
$R_{\mathrm{Jup}}$, and 66 steps in parallax spanning 37.79 to 38.00
milliarcseconds (or kpc$^{-1}$).

\begin{figure*} \includegraphics[width = \linewidth]{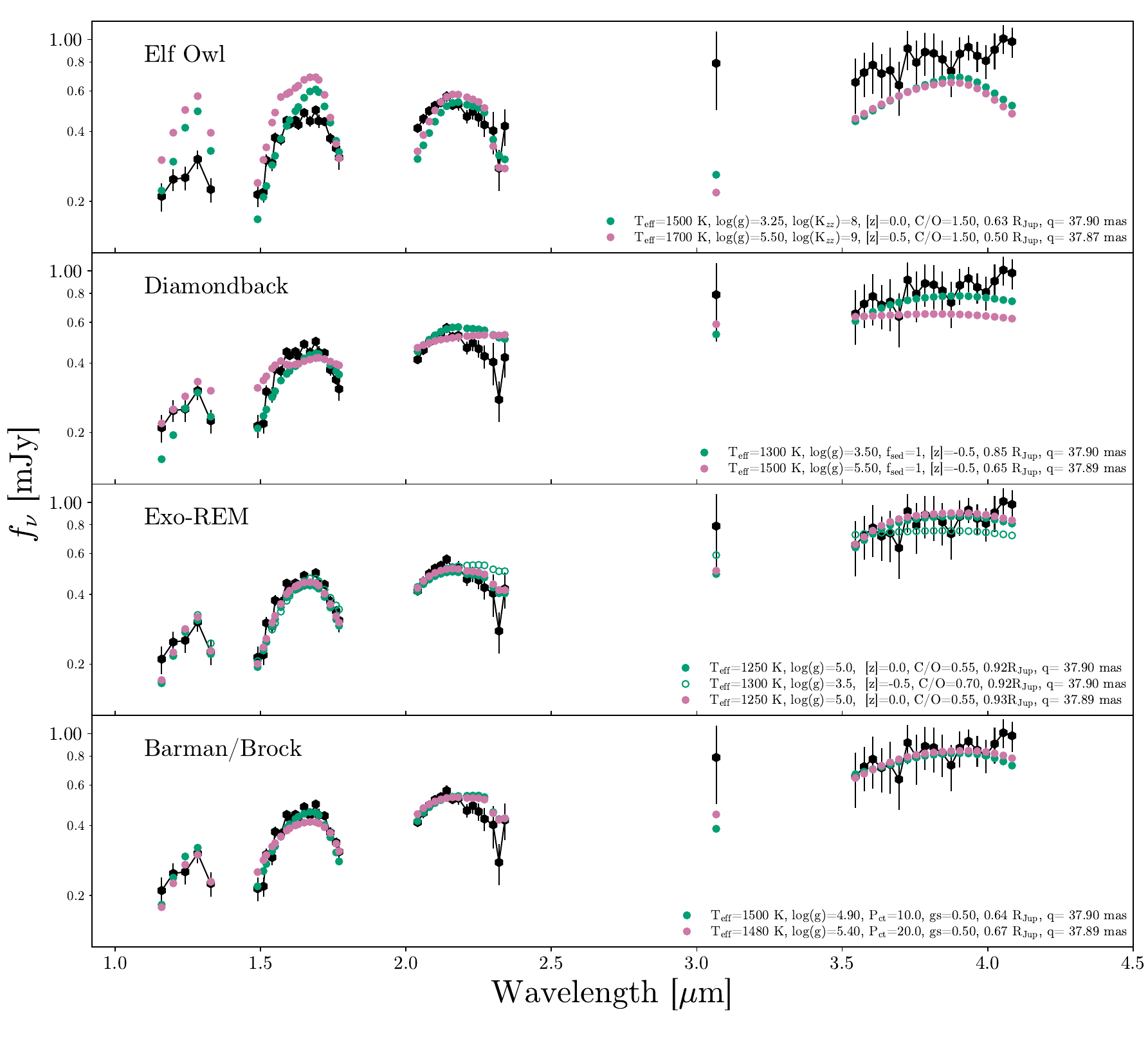}
\caption{Best-fit synthetic atmosphere models from four model grids. In each
case, green markers show fits with a gaussian prior on parallax, a log-flat
prior on radius and surface gravity and a flat prior on effective temperature.
Where applicable, a log-flat prior is used on metallicity and $K_{zz}$ and flat
priors on C/O, and cloud parameters. Pink markers use a joint prior on mass and
surface gravity informed by dynamical mass constraints. See text for details.}
\label{fig:SynthFits} \end{figure*}

We show the best-fitting models from each grid with green markers in the panels
of Figure \ref{fig:SynthFits}. Formal parameter uncertainties for atmospheric
parameters in each case are very tight, not extending beyond one grid point at
the $95\%$ confidence level for all grids except Exo-REM.  For the Exo-REM
grid, there is one additional model within $95\%$ confidence.  In all cases
there is a small uncertainty in radius and parallax, but the precise
\textit{Gaia} parallax measurement is highly constraining.

\begin{figure*} 
\includegraphics[width = \linewidth]{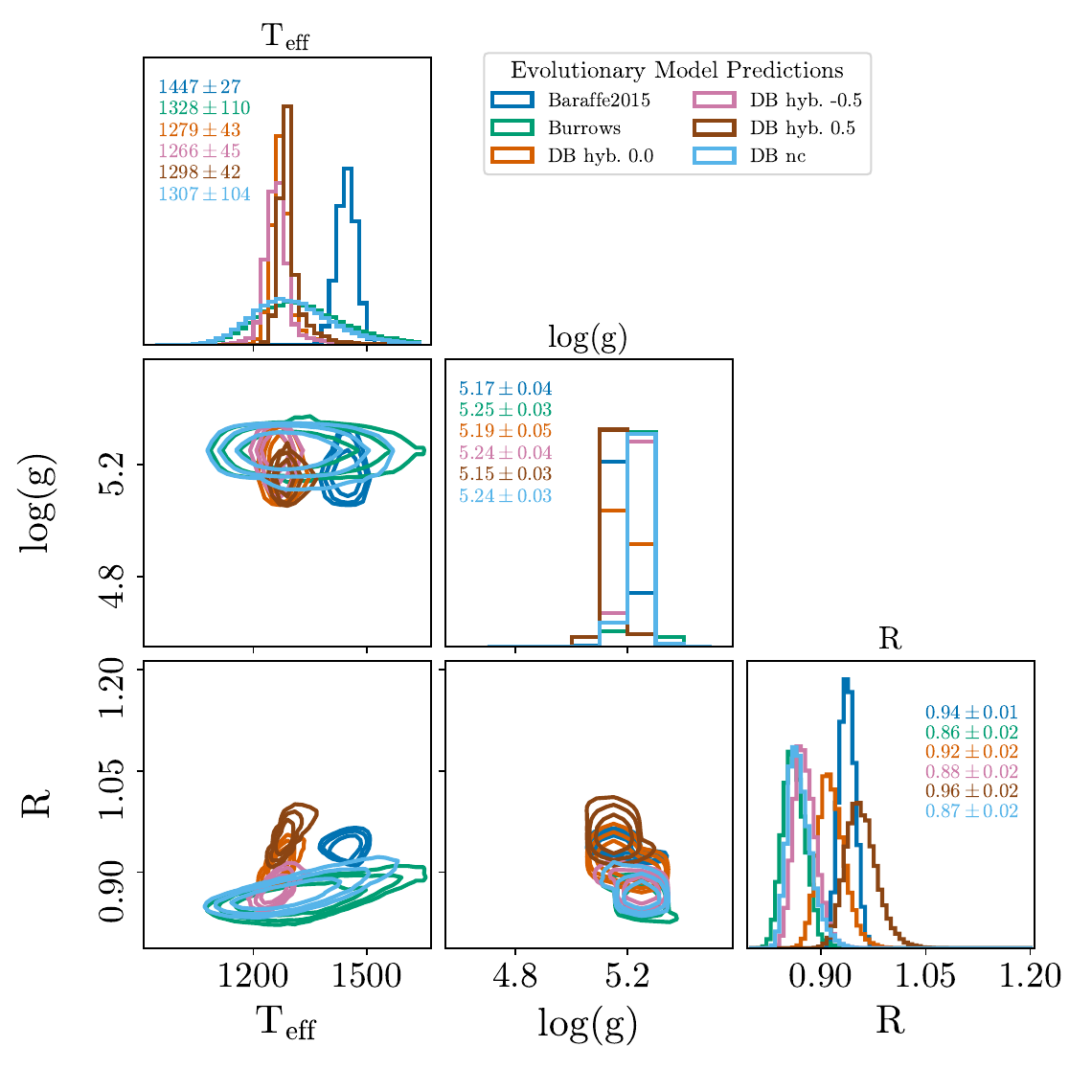} 
\caption{The distribution of evolutionary model predictions for HD~33632~Ab
after propagating uncertainties in the system age and companion mass.}
\label{fig:EvoPredictions} 
\end{figure*}

\section{Discussion} \label{sec:discussion} 
\subsection{Evaluating Fits of Synthetic Spectra}

In Figure \ref{fig:SynthFits} it is clear that the cloudy grids can better
approximate the SED of HD~33632~Ab. This is not surprising given the $\sim L9$
spectral type suggested by our analysis in Section \ref{sec:Templates}.  The
best-fit cloud parameters suggest vertically extended clouds with both
$f_{\mathrm{sed}}$ and $P_{\mathrm{c}}$ at their grid limits for cloud
thickness. These parameter values are somewhat at odds with expectations for an
L9-type object, where the processes of cloud particle precipitation and
dispersal that drive the L/T transition should have begun
\citep{Stephens2009, Brock2021}.  

Across the grids, best-fit effective temperatures range from 1250 K to 1500 K,
best-fit surface gravities range from $\log(g)=3.25 - 5.0$, and best fit
metallicity ranges from $\log(\frac{z}{z_{\odot}})=-0.5 - 0.0$ suggesting
systematic uncertainty much larger than the less-than-one-grid-step formal fit
uncertainty within each grid. 

\subsection{Atmosphere Fits Compared to Evolutionary Models} 
We compare the best-fit atmospheric model parameters presented in the previous
Section to the predictions of several brown dwarf evolution models. For this
purpose, we assume the mass and age constraints are uncorrelated and create
$10^5$ (age, mass) pairs, drawing ages from a normal distribution with mean 1.7
Gyr and a standard deviation of 0.4 Gyr \citep[based on the results
of][]{Brandt2021} and masses from a generalized extreme value distribution
\citep{Possolo2019} representing a mass constraint of
$52.8^{+2.6}_{-2.4}~M_{\mathrm{Jup}}$ \citep{ElMorsy2025}.  We then use SPLAT
\citep{Burgasser2017} to interpolate a variety of evolution models at each of
the age-mass pairs to produce distributions of effective temperature, surface
gravity, and object radius.  

For this work, we manually add the Sonora Diamondback evolution model data into
SPLAT to take advantage of its interpolation architecture.  Results for the
Sonora Diamondback hybrid cloud models at different metallicities as well as
the Diamondback cloud free models are shown in Figure \ref{fig:EvoPredictions}.
The \citet{Burrows2001} and \citet{Baraffe2015} models are also displayed.
While there is some variation in the predictions of the various models,
especially when considering cloudy versus no-cloud models, it is clear that it
is very unlikely for HD~33632~Ab to have a surface gravity below $\log(g)=5.0$
or a radius below $0.75~R_{\mathrm{Jup}}$. Only the higher-gravity Exo-REM
model is consistent with these expectations. For the other grids, the best-fit
atmospheres fall outside this range for at least one parameter.  Consequently,
a sense of unreliability surrounds the interpretation of the model parameters,
especially those that are known to correlate with surface gravity, like
metallicity \citep[e.g.,][]{Skemer2016}.

\subsubsection{Mass Constrained Fits}
In search of more physically plausible atmosphere models, we re-run our fits to
the data using a joint prior on object radius and surface gravity based on the
precise dynamical mass measurement. In this case, we optimize
\begin{equation}
\log{\mathcal{P}} = -0.5\chi^{2} + log(\mathcal{G}) + log(\mathcal{O}),
\end{equation}
where $\chi^2$ and $\mathcal{G}$ are as described above, and $\mathcal{O}$ is
\begin{equation}
\mathcal{O} = 
\frac{1}{\sqrt{2\pi(2.5)^2}}e^{-\frac{(m-52.8~M_{\mathrm{Jup}})^2}{2(2.5~M_{\mathrm{Jup}})^2}},
\end{equation}
with $m=\frac{gR^2}{G}$, $g$ the surface gravity, $R$ the object radius, and
$G$ the gravitational constant. These fits are shown in Figure
\ref{fig:SynthFits}, with pink markers.

This approach alleviates some tension with evolutionary model predictions for
surface gravity in the Sonora models, but in each case tension with
expectations for the object radius are increased. Since we did not interpolate
these grids, the coarse sampling challenges the fitting algorithm as no option
is well suited to satisfy the mass prior with a reasonable radius.

For the interpolated Barman/Brock models, this approach resulted in smaller
changes to best-fit model parameters, partly because the best-fit previously
had a surface gravity more in line with expectations compared to the Sonora
models. The object radius did increase some, but not enough to coincide with
the range of evolutionary model predictions. The $P_{\mathrm{c}}$ parameter
increased, suggesting a deeper cloud and better tuned cloud structure may be
necessary to fit the data. 

The dynamical-mass based prior rules out the low-gravity, low-metallicity, high
C/O Exo-REM model, leaving only the higher-gravity, solar composition model.
This model is nearly a perfect match to the predictions of the evolutionary
models, with the surface gravity, $\log(g)=5$, being slightly too low. However,
the Exo-REM grid does not include models above $\log(g)=5$, so the small, 0.1
dex, discrepancy can be understood as a limitation of the grid range. We note
that our optimal Exo-REM model is the same as that found by \citet{ElMorsy2025}
in their fits to Charis and NIRC2 data alone.

\subsubsection{Investigating Cloud Structure}

\begin{figure*} 
\includegraphics[width = \linewidth]{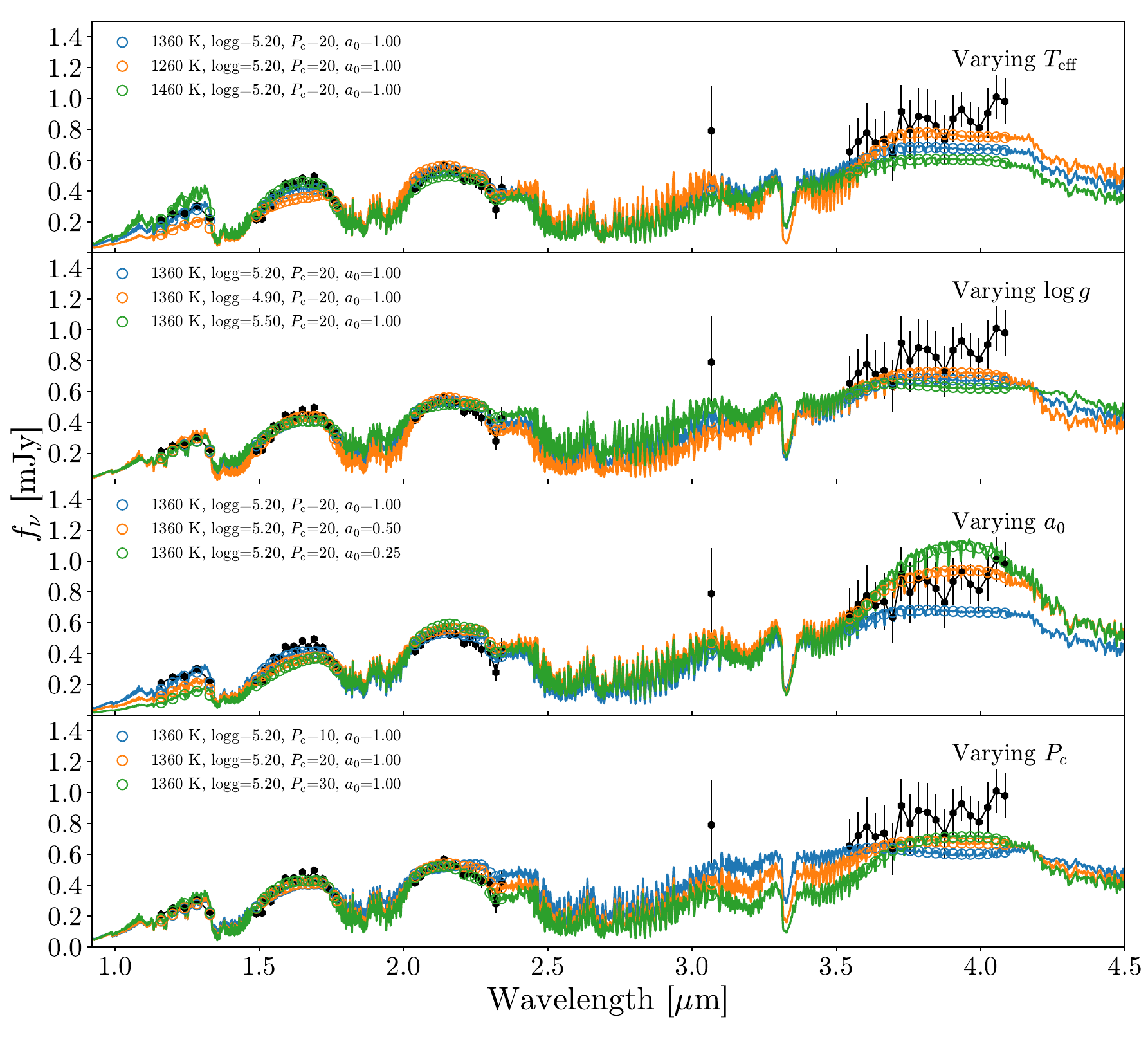} 
\caption{Varying atmosphere model parameters for HD~33632~Ab around the best evolutionary
model-constrained fit from the Barman/Brock grid.}
\label{fig:VaryParams} 
\end{figure*}

Since our interpolated Barman/Brock models include the effective temperature
and surface gravity predicted by evolutionary models, we run an additional
constrained fit for this grid. We again make use of Equation \ref{eqn:Fit} now
using evolutionary model predictions to restrict $T_{\mathrm{eff}}$ to the
range 1100-1600, $\log(g)$ to the range $5.05 -
5.35$, and Radius to the range 0.8 -- 1.02~$R_{\mathrm{Jup}}$. The best-fit
  from this approach is shown in each row of Figure \ref{fig:VaryParams} with
  blue markers.

The constrained fit includes a deeper cloud with larger grains than the
unconstrained fit, consistent with our understanding of the processes driving
the L/T transition \citep{Brock2021}. Overall the model is reasonably close to
the data, however it is too bright in the K-band and too flat in the L-band.

In Figure \ref{fig:VaryParams} we show how the model varies as a function of
the four free parameters, to provide some guidance in interpreting the
remaining tension between the model and the data. We let the overall flux
scaling (that is, the object radius) vary for an optimal fit in each case to
emphasize differences in spectral shape.

Effective temperature is constrained by the near-IR colors, specifically the
J-band flux scaling relative to H and K-band. Colder models begin to exhibit
stronger methane absorption at $\sim3.3~\micron$. Surface gravity changes of
0.3 dex in either direction do not seem to cause large changes in the model
  spectra, outside the unconstrained region between $\sim2.4~\micron
- 3.1~\micron$. The mean cloud particle size changes both the spectral slope
across the L-band and the relative scaling of the J and H bands compared to
K-band. The optimal fit prefers to match the near-IR data at the expense of
a relatively poor fit to the ALES data, owing to the larger number of high
signal-to-noise data points in the near-IR. However, the spectral slope of the
ALES data suggests that a tuned model with smaller cloud particles may be
necessary to provide a good fit.  Varying the cloud top-pressure results in
significant variation in the model spectra, especially at the wavelengths
corresponding to the P-, Q-, and R-branch ro-vibrational transitions of
methane, $\sim3.1~\micron - 3.7~\micron$. Our inability to detect HD~33632~Ab
between $\sim3.1~\micron - 3.5~\micron$, may suggest low intrinsic flux through
this range, possibly due to deeper cloud structures. Evidently, constraining
the architecture of clouds in the atmosphere of HD~33632~Ab would greatly
benefit from measurements in this range. ALES observations under better
telluric throughput conditions than presented here could contribute meaningful
constraints in this range. 

\subsection{Comparison to HR 8799 c, d, e}

\begin{figure*} 
\includegraphics[width = \linewidth]{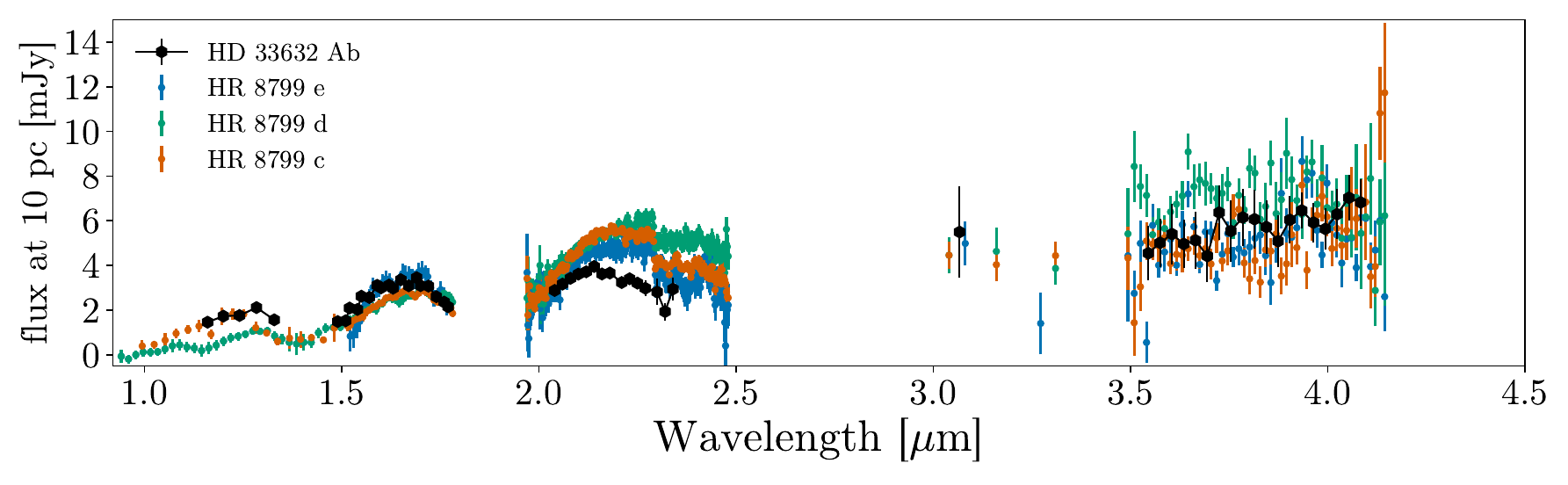} 
\caption{HD 33632 Ab compared to the SEDs of HR 8799 c, d, e adopted from
\citet{Doelman2022} and using the K-band measurements from \citet{Nasedkin2024}.}
\label{fig:HR8799} 
\end{figure*}

HD 33632 Ab exhibits spectral similarity to the planets in the HR 8799
system, which also have spectral types at the L/T transition
\citep[e.g.,][]{Greenbaum2018}. In Figure \ref{fig:HR8799}, we compare the
$1-4.1~\micron$ spectrum of HD~33632~Ab to the spectra of HR~8799~c, d, and
e \citep[including data from][]{Doelman2022, Galicher2011,Skemer2014,
Zurlo2016,Oppenheimer2013, Greenbaum2018, Nasedkin2024}. All spectra have been scaled to
a distance of 10 pc to facilitate comparison.

Spectral analysis suggests that HD~33632~Ab has an effective temperature
$\sim200~K$ warmer than the three HR~8799 planets considered here
\citep{Doelman2022}. Consequently, the close overlap in general flux level
between the planets and HD~33632~Ab provides evidence that the younger planets
have larger radii, consistent with our expectations from evolutionary models.

In the K-band, major opacity sources include collisionally induced
absorption of molecular hydrogen, and ro-vibrational bands of both methane and
carbon monoxide. Combined, these opacity sources result in K-band flux being
highly-sensitive to photospheric pressure, which is effected by surface
gravity, metallicity, and clouds. The lower K-band flux of HD~33632~Ab compared
to the planets, must in part be due to significantly higher surface gravity in
the more evolved (smaller radius) and more massive object.

HR~8799~d has a lower J-band flux and a flatter L-band spectral slope than
HD~33632~Ab. Figure \ref{fig:VaryParams} shows that J-band flux relative to the
other bands is affected by changes in temperature and cloud particle size. We
also see that L-band spectral slope is affected by cloud particle size.
HR~8799~d and HD~33632~Ab likely have different temperatures and cloud
structures.

\section{Conclusion}\label{sec:conclusion} 

We present high-contrast ALES integral field spectroscopy observations of the
HD~33632 system, resulting in a flux calibrated $3-4.12~\mu\mathrm{m}$ spectrum
of the brown dwarf companion HD~33632~Ab. We compile a spectral energy
distribution for the brown dwarf combining Subaru/Charis data with our new
LBTI/ALES data. The low-resolution $1-4.12~\mu\mathrm{m}$ spectrum is matched
to late L-dwarf template spectra with best fits found for field-age
L7--L9 type objects (Figure \ref{fig:TemplatePlot}).

Synthetic atmosphere models with clouds provide a reasonably close match to the
data (Figure \ref{fig:SynthFits}), however, the cloud-free Elf Owl models do
not reproduce the shape of the SED. This is consistent with clouds playing
a central role in shaping the appearance of L-dwarf spectra.

Dynamical constraints from radial velocity monitoring and astrometric
measurements of the system yield a tight mass constraint on HD~33632~Ab
\citep{ElMorsy2025}. Combining mass with age estimates for the system enables
the use of brown dwarf evolutionary models to predict the bulk physical
properties of HD~33632~Ab.  Various families of evolutionary models each
including different treatment of clouds and different bulk metallicities all
suggest a tight distribution of surface gravities and object radii when
uncertainty in mass and age are propagated (Figure \ref{fig:EvoPredictions}).
Our best-fit model atmospheres imply significant tension with the evolutionary
model consensus, suggesting unphysical best-fit parameters and creating some
uncertainty in the interpretation of parameters like metallicity and C/O ratio.

We demonstrate an atmosphere model fitting approach using a joint prior on mass
and gravity based on dynamical constraints. This approach emphasized the need
for finely sampled grids when using such precise mass measurements. Application
of mass-based priors to our interpolated grid of Barman/Brock models did
alleviate some of the tension with evolutionary models, but the best fit radius
is still $\sim30\%$ too small. For the Exo-REM grid, the mass-based prior rules
out fits incompatible with evolutionary model predictions.

We run a third optimization approach using the predictions of
evolutionary models directly as priors in our atmosphere fit. Given the tight
evolutionary model constraints we only apply this to the interpolated
Barman/Brock models. This exercise identifies a deeper cloud with larger grains
than the unconstrained fit, consistent with expectations for the L/T
transition \citep{Brock2021}.

Finally, we compare to HR~8799~c, d, and e, showing general spectroscopic
similarities between the L/T transition objects, but with significant
differences likely due to differences in temperature, surface gravity, and
cloud structure.

An on-going LBTI/ALES observational program and an independent JWST Cycle
4 NIRSpec IFU program both aim to collect thermal-IR spectroscopy for an emerging
  population of directly imaged companions whose association with a more
massive host star facilitate dynamical mass constraints from \textit{Gaia} and
compositional constraints from stellar spectroscopic analysis. The goal of both
of these efforts is to create a spectroscopic library of objects with
constrained physical parameters to facilitate progress in understanding the
complex physical processes sculpting the dramatic L/T transition and cloud
processes in general.

\appendix
\section{Signal-to-noise Maps for Selected Wavelength Slices}

\begin{figure*} \includegraphics[width = \linewidth]{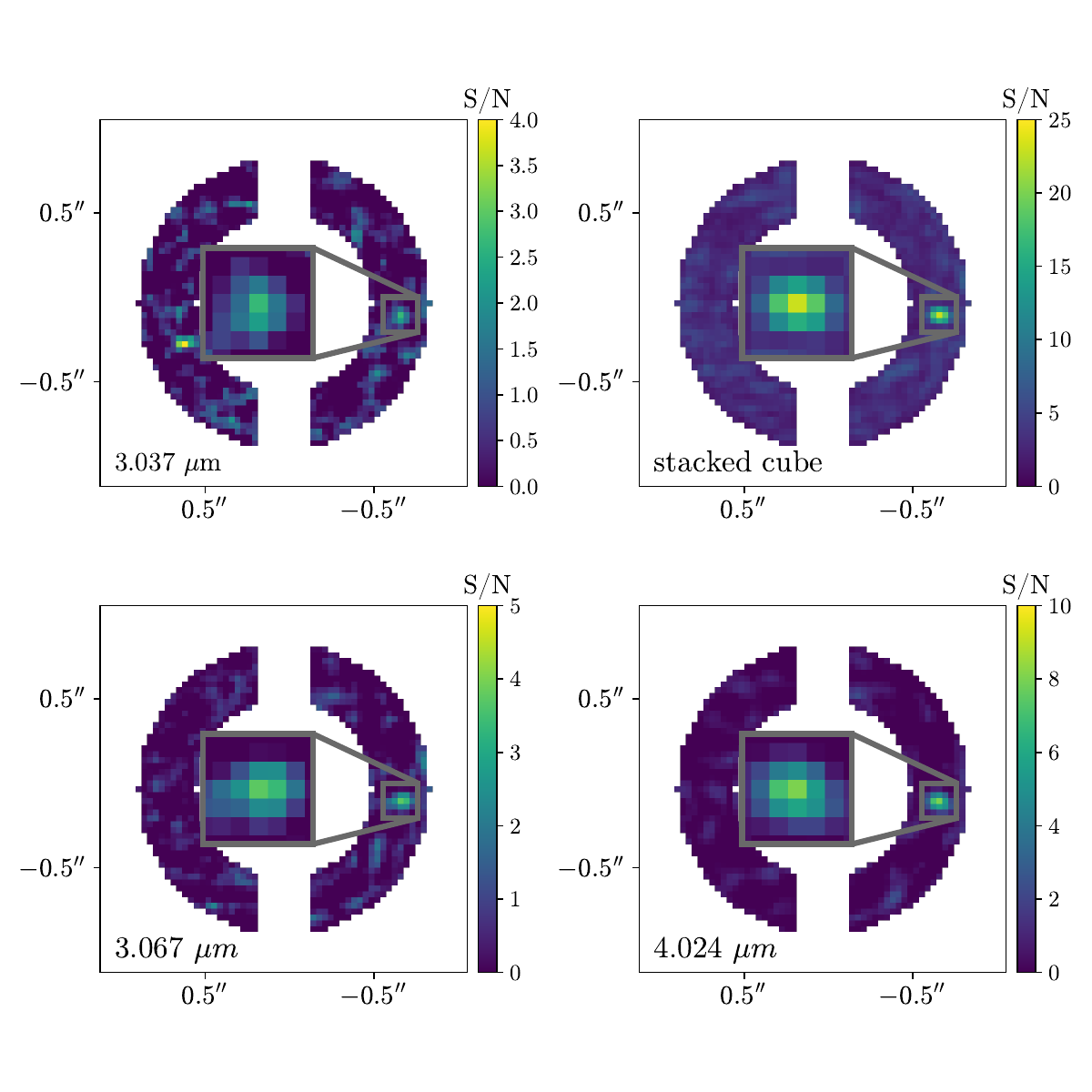}
\caption{Signal-to-noise maps for three wavelength slices and the median
combined spectral cube for our 2021-02-28 ALES observations of HD~33632~Ab. The
inset for each panel highlights the optimization region used for extracting
companion photometry using an iterative negative-planet injection strategy (see
Section \ref{sec:spectralExtraction})} \label{fig:SNImage} \end{figure*}

In Figure \ref{fig:SNImage}, we show signal-to-noise (S/N) maps for selected
wavelength slices and our median collapsed cube.  For this purpose, the signal
in each pixel is taken as the PSF-weighted mean flux in an aperture centered on
that pixel with radius $\lambda/D$, and noise as the standard deviation of flux
measurements in non-overlapping apertures at the same projected separation.

Detector crosstalk results in pixels aligned with the host star
exhibiting complex structure in ALES cubes\citep{Doelman2022}. In our S/N maps,
we mask the regions to the North and South most affected by crosstalk. An inset
image in each panel of Figure \ref{fig:SNImage} highlights the optimization
region used in our negative-planet-injection approach to companion photometry.

\begin{acknowledgements} We thank the LBT director for allocating director's
discretionary time to this project. Thanks to Ilya Ilyin and Jennifer Power for
their support collecting data for this program.

This work benefited from the Exoplanet Summer Program in
the Other Worlds Laboratory (OWL) at the University of California, Santa Cruz,
a program funded by the Heising-Simons Foundation.  

Support for basic research at the U.S. Naval Research Laboratory comes from the
6.1 base program. This paper is based on work funded by NSF Grants 1608834,
1614320 and 1614492.

The LBT is an international collaboration among institutions in the United
States, Italy and Germany.  LBT Corporation partners ar: The University of
Arizona on behalf of the Arizona university system; Istituto Nazionale di
Astrofisica, Italy; LBT Beteiligungsgesellschaft, Germany, representing the
Max-Planck Society, the Astrophysical Institute Potsdam, and Heidelberg
University; The Ohio State University, and The Research Corporation, on behalf
of The University of Notre Dame, University of Minnesota, and University of
Virginia.  We thank all LBTI team members for their efforts that enabled this
work.    
\end{acknowledgements}

\facilities{LBT (LBTI/LMIRCam, LBTI/ALES)}

\software{Astropy \citep{astropy2013},
Matplotlib \citep{matplotlib},
Scipy \citep{2020SciPy-NMeth},
SPLAT \citep{Burgasser2017},
}

\bibliography{ms.bib}
\bibliographystyle{aasjournalv7}
\end{document}